\documentclass[11pt,a4paper]{article}
\usepackage[T1]{fontenc}
\usepackage[utf8]{inputenc}
\usepackage{lmodern}
\usepackage[margin=1in]{geometry}
\usepackage{amsmath,amssymb}
\usepackage{graphicx,booktabs,longtable,array}
\usepackage{microtype,xurl}
\usepackage[font=small,labelfont=bf]{caption}
\usepackage{xcolor}
\definecolor{linkblue}{RGB}{25,65,95}
\usepackage[colorlinks=true,linkcolor=linkblue,citecolor=linkblue,urlcolor=linkblue,
  pdfauthor={Anas H. Alzahrani},
  pdftitle={Communication-constrained multiagent risk estimation: a simulation benchmark for adult sore throat}]{hyperref}
\date{}
\title{\LARGE\bfseries Communication-constrained multiagent risk estimation: a simulation benchmark for adult sore throat}
\author{Anas H. Alzahrani, MD, MPH, PhD\\[0.35em]
\small Department of Preventive Medicine and Public Health\\
\small Faculty of Medicine, King Abdulaziz University\\
\small Jeddah, Saudi Arabia\\[0.35em]
\small \href{mailto:ahalzhrani@kau.edu.sa}{ahalzhrani@kau.edu.sa}\\
\small ORCID: \href{https://orcid.org/0000-0002-1394-9157}{0000-0002-1394-9157}}
\begin{document}\maketitle
\begin{abstract}Distributed predictive agents may need to approach a shared estimate under limited, unreliable communication. We evaluated classical symmetric pairwise gossip among five agents initialized from frozen calibrated predictions in a synthetic adult sore-throat benchmark. Ninety runs across three fixed generator families supplied 270,000 underlying test encounters, each observed through four missingness views. Complete, ring and star graphs were compared at eight interaction budgets and transaction failure probabilities of 0, 0.2 and 0.5. The coordination extension was designed after inspection of the predictive benchmark and is exploratory. Outcomes were disagreement around the conserved arithmetic mean and its exact decomposition of mean-agent Brier error. In the base view at eight attempted pairs and 0.2 failure, disagreement variance was 0.000109 for the complete graph, 0.000134 for the ring and 0.000175 for the star. The paired complete-minus-ring difference was -0.000024 (95\% Monte Carlo interval -0.000030 to -0.000019). The centralized mean had base Brier 0.069781, compared with 0.069368 for Kumo alone. Agreement therefore does not establish superiority to the strongest individual predictor. Code, fixed probability inputs and results support replay without model refitting. The study quantifies finite-budget coordination under specified simulation assumptions; it proposes no new gossip algorithm and establishes no clinical accuracy or utility.\end{abstract}

\section{Introduction}
Distributed agents may hold different probability estimates for the same event, yet have limited opportunities to exchange them. Even when a common aggregate is well defined, the estimates available locally depend on who communicates, whether an exchange succeeds, and when communication stops. Classical opinion averaging and randomized gossip provide a precise framework for examining these effects. \cite{degroot,boyd} A finite communication budget is particularly relevant when an agent must act on its current local state rather than wait for exact agreement.

We study this coordination problem in a fully synthetic adult sore-throat application. Five agents begin with calibrated risk estimates produced by different locally executable prediction pipelines. Each agent stores one estimate per encounter and applies a local update after a successful exchange with a neighbor. The models provide heterogeneous initial opinions; the multiagent mechanism is the subsequent stateful exchange process. The agents are numerical processes, without language-model reasoning, clinical roles, independent patient observations or task planning.

The application concerns whether group A Streptococcus (GAS) is the simulated cause of an adult sore-throat episode within the broader setting of upper respiratory tract infection. It does not cover every respiratory syndrome. Clinical guidance distinguishes risk estimation, testing and treatment, and symptoms alone do not establish etiology. \cite{ref1,ref4} Accordingly, this study evaluates agreement and probability error in specified hypothetical worlds, without interpreting the agents as clinicians or proposing a prescribing pathway.

The contribution is an applied, reproducible benchmark of complete, ring and star communication graphs under finite interaction budgets and atomic transaction loss. We ask how quickly local opinions approach the centralized arithmetic mean, how missing observations change initial disagreement, and how agreement relates algebraically to mean local predictive error. The gossip algorithm is established; no new consensus method is proposed. The coordination study is an exploratory extension designed after inspection of a completed predictive benchmark. That benchmark remains documented as the source of the frozen opinions and as a supporting comparison, rather than being retrospectively described as a multiagent experiment.

\section{Methods}
\subsection{Simulation substrate and analysis chronology}
Three fixed hypothetical data-generating families represented additive, interaction and latent disease/subtype mechanisms. Each family contributed 30 independent runs with seeds 101--130. Within each run, models used 2,000 training and 1,000 separate calibration encounters. Calibrated probabilities were frozen before the communication experiment; no model, calibrator or aggregation weight was fitted or selected using the communication outcomes. The original benchmark compared 12 predictive methods across seven test conditions and included a separate assay-label experiment. Its original designated contrast, graph-summary versus flat-knowledge XGBoost, is retained in Appendix~\ref{app:benchmark}.

The communication extension uses four observation views: base, and 30\% eligible-field missingness under missing completely at random (MCAR), missing at random (MAR), and missing not at random (MNAR). Each run contributes 3,000 test encounters per view. The four views share the same underlying cases and outcomes: 90 runs therefore supply 270,000 distinct underlying test encounters and 1,080,000 encounter views, not 1,080,000 independent patients. Base missingness is 10\% in nine eligible fields; age, duration and cough remain observed. The corresponding expected proportions over all twelve fields are approximately 7.5\% and 22.5\%. The numerical mechanisms and full predictive benchmark are specified in the appendices.

The extension design, seed streams and designated exploratory contrast were recorded locally before its outcomes were computed, but after the predictive results had been inspected. Selection of the five pipelines and the four observation views was therefore post hoc. This record establishes computational chronology, not prospective or external preregistration. All communication analyses and intervals are exploratory, without multiplicity adjustment.

\subsection{Agents, states and initial opinions}
For encounter $c$, agent $i\in\{1,\ldots,5\}$ starts at $p_{ci}^{(0)}\in[0,1]$, the saved sigmoid-calibrated probability from one of five pipelines: regularized logistic regression, additive explainable boosting (EBM), Kumo Tabular Small, flat-knowledge XGBoost, and graph-summary XGBoost. These five pipelines span statistical, additive, tree and pretrained tabular approaches already evaluated on identical data. They are not independent experts: their training samples, target and observed encounter features overlap. Differences in feature representation and preprocessing remain part of their initial opinions.

A seed-fixed random permutation assigns the five predictors to nodes once per run, shared across observation views, communication graphs and failure settings. Thus the star hub is not deliberately assigned the best predictor. Each simulated encounter has its own sequence of selected edges and transaction outcomes. The implementation batches these independent encounter-level trajectories for computation; it does not represent an actual distributed network deployment.

The communication graph is distinct from the clinical feature graph used inside one XGBoost pipeline. The latter constructs fixed feature summaries before prediction; it neither chooses communication partners nor changes during gossip. No prediction model is retrained during interaction.

\subsection{Communication protocol and cost}
We compare an undirected complete graph $K_5$ with ten edges, a ring $C_5$ with five edges, and a star with four edges. At attempted interaction $t$, an undirected edge $(i,j)$ is sampled uniformly from the chosen graph. With transaction failure probability $q\in\{0,0.2,0.5\}$, both endpoint states remain unchanged. Otherwise, an atomic symmetric exchange sets
\[
 p_{ci}^{(t+1)}=p_{cj}^{(t+1)}=
 \frac{p_{ci}^{(t)}+p_{cj}^{(t)}}{2},
\]
while every other state remains unchanged. Both reciprocal updates commit together or neither commits. This is a whole-transaction failure model, not independent packet loss, asynchronous delayed delivery, or a Byzantine-fault model. Uniform edge sampling is supplied by the simulation scheduler; the payload update requires only the two endpoint states. \cite{boyd}

Nested trajectory checkpoints use $B\in\{0,1,2,4,8,16,32,64\}$ attempted pair interactions per encounter. Each attempted interaction costs two scalar probability transmissions, including failed attempts; delivered payloads equal twice the successful transactions. Thus the maximum budget is 128 attempted scalar payloads per encounter. This abstract cost excludes edge scheduling, addressing, handshakes, acknowledgements, latency and packet headers. It is not a measured byte count or wall-clock communication saving. The centralized arithmetic mean is an ideal full-information reference whose communication cost is not modeled.

Within a run and encounter, common random uniforms are reused across graph, loss and observation-view comparisons; each graph maps the edge-selection uniform to its own edge list. Loss uniforms are shared, and larger budgets continue the same trajectory. Comparisons therefore use paired starting opinions and random inputs. Equal total interaction budgets do not imply equal per-node participation: every complete or ring node is an endpoint with probability $2/5$ per attempt, whereas the star hub participates in every attempt and each leaf with probability $1/4$.

\subsection{Coordination outcomes and predictive-error decomposition}
The conserved full-information mean for encounter $c$ is
\[
 \bar p_c=\frac{1}{5}\sum_{i=1}^{5}p_{ci}^{(0)}.
\]
For $N=3000$ encounters in a run/view, disagreement variance and its root-mean-square (RMS) residual are
\[
 D_B=\frac{1}{5N}\sum_{c=1}^{N}\sum_{i=1}^{5}
       (p_{ci}^{(B)}-\bar p_c)^2,\qquad R_B=\sqrt{D_B}.
\]
These are the principal coordination outcomes. Reported RMS summaries average run-specific square roots, rather than taking the square root of an aggregated variance. Let $Y_c$ be simulated disease truth. Define mean-agent Brier and centralized-mean Brier by
\[
 A_B=\frac{1}{5N}\sum_{c,i}(p_{ci}^{(B)}-Y_c)^2,
 \qquad C=\frac{1}{N}\sum_c(\bar p_c-Y_c)^2.
\]
Mean conservation gives the exact decomposition
\begin{equation}\label{eq:decomposition}
 A_B=C+D_B.
\end{equation}
The cross term vanishes because local deviations sum to zero. A successful averaging step reduces the sum of squared local deviations for that encounter by $(p_{ci}-p_{cj})^2/2$; a failed step changes nothing. Therefore $D_B$ and $A_B$ cannot increase along a realized trajectory. These are mathematical properties of symmetric averaging, not empirical discoveries or independent validation signals. Between graph or loss settings with the same initial opinions, differences in $A_B$ equal differences in $D_B$; outcome labels cancel.

Mean-agent Brier averages errors of five local states. It is not the error of a single chosen team output, and does not imply that every predictor improves when moved toward the pool. We retain each named agent's Brier score, the initial individual predictors, the no-communication state, successful exchange counts, node participation and the maximum casewise mean-conservation residual. Simulated labels enter evaluation only; they do not determine messages, partner choice, random mapping or update rules.

\subsection{Contrasts, uncertainty and verification}
The designated exploratory contrast is complete-minus-ring disagreement variance in the base view after eight attempted pairs (16 attempted scalar payloads), with transaction failure probability 0.2. All other budgets, graph comparisons and observation-view contrasts are descriptive extensions. Each of the three fixed families receives equal weight. Within-family summaries use 30 runs; paired contrasts use within-run differences. With $v_f=s_f^2/(30\times3^2)$, Monte Carlo standard error is $\sqrt{\sum_f v_f}$ and pointwise 95\% intervals use a Student $t$ quantile with degrees of freedom $(\sum_f v_f)^2/\sum_f[v_f^2/29]$. Intervals describe finite simulation repetition conditional on the selected families and reference draws, not clinical generalization. Nodes, encounters and repeated observation views are not treated as independent replicates for these intervals.

Saved probabilities were joined by encounter identifier, with outcome agreement and source-file hashes checked. Compact float64 inputs, deterministic random seeds, source, configuration and results accompany the paper so the communication experiment can run without refitting the component models. Numerical checks cover scalar-versus-vector updates, probability bounds, mean conservation, the Brier decomposition, nonincreasing disagreement, all-failure no-ops, expected output counts and agreement of initial Brier scores with the frozen component benchmark. A transfer-inclusive NumPy/CuPy timing probe informed backend selection. These checks establish arithmetic and provenance, rather than realism of the clinical or network assumptions.
\section{Results}\subsection{Executed trajectories and initial opinions}
All 360 run/\allowbreak{}view input matrices and 1,800 initial model Brier scores passed alignment and reconciliation checks. The experiment produced 25,920 run--view--graph--loss--budget summaries and 2,520 baseline records. Repeated zero-budget and centralized-reference values are retained for transparent indexing; they are not distinct independent observations. Figure 1 specifies the interaction process, Figure 2 shows convergence and predictive-error decomposition, and Figure 3 compares observation views.

In the base view, mean no-communication agent Brier was 0.070432 and centralized-mean Brier was 0.069781. Initial disagreement variance was 0.000651, accounting for their difference. The five individual Briers were 0.069900 for logistic regression, 0.070156 for EBM, 0.069368 for Kumo, 0.071731 for flat XGBoost and 0.071003 for graph-summary XGBoost. These initial errors describe frozen component opinions, not communication outcomes.

\subsection{Designated exploratory coordination contrast}
At eight attempted pairs and transaction failure probability 0.2, complete-graph disagreement variance was 0.000109, compared with 0.000134 for the ring and 0.000175 for the star. The paired complete-minus-ring difference was -0.000024 (95\% Monte Carlo interval -0.000030 to -0.000019), an 18.3\% reduction relative to the ring mean. This percentage is a ratio of unrounded means, without a separately estimated interval. Family-specific paired differences were additive: -0.000012, interactions: -0.000026, latent mixture: -0.000035.

The corresponding mean-agent Briers were 0.069890, 0.069914 and 0.069955. Their complete-minus-ring contrast equals the disagreement contrast by construction; it is not a second independent efficacy finding. Mean successful exchanges were 6.397 per encounter, corresponding to 12.795 delivered scalar payloads out of 16 attempted. All graphs share the realized transaction outcomes.

\begingroup\small
\begin{longtable}{@{}>{\raggedright\arraybackslash}p{0.1600\linewidth}>{\raggedright\arraybackslash}p{0.1600\linewidth}>{\raggedright\arraybackslash}p{0.1800\linewidth}>{\raggedright\arraybackslash}p{0.2100\linewidth}>{\raggedright\arraybackslash}p{0.1900\linewidth}@{}}
\caption{Finite-budget coordination in the base observation view}\label{tab:coordination}\\
\toprule
\textbf{Graph} & \textbf{Failure probability} & \textbf{RMS residual} & \textbf{Disagreement variance} & \textbf{Mean-agent Brier} \\
\midrule\endfirsthead
\caption[]{Finite-budget coordination in the base observation view (continued)}\\
\toprule
\textbf{Graph} & \textbf{Failure probability} & \textbf{RMS residual} & \textbf{Disagreement variance} & \textbf{Mean-agent Brier} \\
\midrule\endhead
\midrule\multicolumn{5}{r}{Continued on next page}\\\endfoot
\bottomrule\endlastfoot
Complete & 0.0 & 0.007934 & 0.000065 & 0.069846 \\
Ring & 0.0 & 0.009237 & 0.000089 & 0.069870 \\
Star & 0.0 & 0.010959 & 0.000125 & 0.069905 \\
Complete & 0.2 & 0.010271 & 0.000109 & 0.069890 \\
Ring & 0.2 & 0.011308 & 0.000134 & 0.069914 \\
Star & 0.2 & 0.012977 & 0.000175 & 0.069955 \\
Complete & 0.5 & 0.014668 & 0.000223 & 0.070004 \\
Ring & 0.5 & 0.015119 & 0.000238 & 0.070019 \\
Star & 0.5 & 0.016450 & 0.000281 & 0.070061 \\
\end{longtable}
\noindent All rows use eight attempted pairs (16 attempted scalar payloads) per encounter. Each value averages 90 runs with equal fixed-family weights. RMS averages run-specific square roots. Mean-agent Brier equals centralized-mean Brier plus disagreement variance; it does not describe a selected team output. Pointwise conditional Monte Carlo intervals for all cells are supplied in the ancillary results.
\par\endgroup
\subsection{Budgets, missingness and individual-agent effects}
Disagreement and mean-agent Brier were nonincreasing along every realized trajectory within numerical tolerance, as expected from symmetric averaging. At 64 attempts, mean RMS residuals at failure probability 0.2 were complete: 0.00001919, ring: 0.00008206, star: 0.00018217. Full trajectories and pointwise Monte Carlo intervals are retained for every graph, loss level and observation view; 64 attempts should not be interpreted as exact consensus.

Initial disagreement variances for base, MCAR, MAR and MNAR views were 0.000651, 0.000986, 0.000832, 0.000718. Under the designated eight-attempt, 0.2-failure setting, complete-graph residual variances were 0.000109, 0.000165, 0.000139, 0.000120, respectively. The observation views are paired within the same underlying encounters. Differences in agreement reflect the component opinions induced by those views, without establishing clinical robustness.

The centralized pool had Brier 0.069781 compared with 0.069368 for Kumo alone, an absolute difference of 0.000413. Thus the ideal consensus target did not outperform the strongest initial component in this base benchmark. Kumo-associated local Brier at the designated complete-graph checkpoint was 0.069716, while mean-agent Brier decreased. Table 2 reports each named component before and after communication, illustrating why average local improvement must be distinguished from benefit for each agent.

\begingroup\small
\begin{longtable}{@{}>{\raggedright\arraybackslash}p{0.3000\linewidth}>{\raggedright\arraybackslash}p{0.1500\linewidth}>{\raggedright\arraybackslash}p{0.1500\linewidth}>{\raggedright\arraybackslash}p{0.1500\linewidth}>{\raggedright\arraybackslash}p{0.1500\linewidth}@{}}
\caption{Individual-agent Brier before and after communication}\label{tab:individual}\\
\toprule
\textbf{Initial predictor} & \textbf{No communication} & \textbf{Complete, 8 pairs} & \textbf{Ring, 8 pairs} & \textbf{Star, 8 pairs} \\
\midrule\endfirsthead
\caption[]{Individual-agent Brier before and after communication (continued)}\\
\toprule
\textbf{Initial predictor} & \textbf{No communication} & \textbf{Complete, 8 pairs} & \textbf{Ring, 8 pairs} & \textbf{Star, 8 pairs} \\
\midrule\endhead
\midrule\multicolumn{5}{r}{Continued on next page}\\\endfoot
\bottomrule\endlastfoot
Logistic regression & 0.069900 & 0.069813 & 0.069804 & 0.069763 \\
Explainable boosting & 0.070156 & 0.069885 & 0.069850 & 0.069893 \\
Kumo Tabular Small & 0.069368 & 0.069716 & 0.069713 & 0.069611 \\
Flat XGBoost & 0.071731 & 0.070068 & 0.070184 & 0.070417 \\
Graph-summary XGBoost & 0.071003 & 0.069966 & 0.070021 & 0.070093 \\
\end{longtable}
\noindent Base view, transaction failure probability 0.2; means over 90 runs with equal fixed-family weights. Agent names identify the predictor that supplied the initial state, even after that state has been mixed with neighbors. Centralized-mean Brier is 0.069781. This ideal reference has no modeled communication cost.
\par\endgroup
\subsection{Numerical verification and computation}
Maximum mean-conservation residual was 5.55e-16; maximum Brier-decomposition discrepancy was 2.70e-16. The maximum positive per-case variance increment was 1.63e-19, consistent with roundoff. The matched CPU probe took a median 0.200 seconds per four-view run bundle. CUDA could not be timed because CuPy reported out-of-memory during warm-up; CPU was used without a claim that it was faster than GPU execution. The complete extension executed using numpy in 24.1 seconds, including saved-result construction. Timings reflect the available shared workstation and are not network-deployment measurements. The ancillary verification records contain the scalar-reference, all-failure, input-reconciliation and CPU timing and unsuccessful CUDA probe records.

\section{Discussion}
The experiment isolates a narrow but explicit multiagent problem: different local predictive opinions must approach a common mean through limited, unreliable pairwise exchanges. Graph structure changes how quickly information spreads under an equal total attempt budget. Transaction failure reduces the number of successful updates. The resulting residual differences quantify finite-budget agreement for these initialized networks, rather than demonstrating a new consensus algorithm. \cite{boyd}

The predictive interpretation follows from Equation~\ref{eq:decomposition}. Averaging cannot increase disagreement or mean-agent Brier, while the centralized-mean Brier remains fixed. It cannot establish that the target itself is accurate, calibrated, preferable to the strongest individual predictor, or clinically useful. Shared model biases are conserved in that target. In particular, the five agents are trained on overlapping synthetic data; agreement should not be interpreted as independent corroboration. The individual-agent comparisons distinguish a reduction in average local error from improvement for each component.

The controlled setup deliberately omits several properties of deployed agent systems. There are only five homogeneous scalar-state communication nodes; model training is centralized and completed beforehand. Agents do not collect new evidence, optimize separate objectives, allocate tasks, negotiate, explain decisions, or execute language-model reasoning. Graphs are fixed and connected, edge scheduling is external, transactions are instantaneous and atomic, and there are no delays, correlated outages, malicious nodes, resource heterogeneity or stopping decisions driven by observed disagreement. The experiment does not compare learned communication, push-sum, event-triggered consensus or adaptive agent selection. Those would require separate implementations and budgets. A star's concentrated node participation also means equal total payload cost is not equal per-agent workload or resilience.

The application adds further limits. All mechanisms are investigator-specified, without fitting to an empirical Saudi or other clinical joint distribution. The generator families, missingness rates, single training size and equal family weights define a restricted benchmark. Clinical-score mappings and graph groups are constructed under stated operationalizations; one degree-preserving graph shuffle is insufficient to establish the general value of clinical graph structure. Kumo and EBM use native missing-value handling while other pipelines use imputation, so differences reflect complete pipelines. No external clinical validation, clinician assessment, treatment policy, explanation fidelity, antibiotic consumption or patient benefit was evaluated.

The supporting assay experiment illustrates a separate issue that communication does not solve: a model trained on assay positivity can target a different quantity from etiologic disease when carriage, assay error and selective verification intervene. \cite{ref2} We did not rerun the five-agent experiment under those altered training targets. Nor did we evaluate the prevalence and presentation-shift views in the communication extension. Future studies could vary network size, correlated failure, heterogeneous local evidence, alternative aggregation targets and observation mechanisms. Such extensions should be specified before inspecting their outcomes. Patient-facing work would require a separate protocol, clinical review, appropriate institutional determinations and untouched validation data.

\section{Conclusion}
This study provides a reproducible simulation of finite-budget coordination among five predictor-backed agents. Complete, ring and star networks are compared under atomic transaction loss using frozen opinions from a synthetic adult sore-throat benchmark. Agreement and mean-agent Brier are linked by an exact variance decomposition, so improved agreement must not be interpreted as independent evidence of improved clinical reasoning. The findings concern the specified communication and data-generating mechanisms, without establishing clinical readiness.
\section*{Declarations}
\subsection*{Ethics and consent}
This computational study used only simulated data and involved no patient records or human participants. No institutional ethics approval or formal exemption is claimed.

\subsection*{Funding}
This work was supported by the Saudi National Institute of Health (Saudi NIH) under Grant PRI01-2401-KAU19-46641119.

\subsection*{Competing interests}
The author declares no competing interests relevant to this work.

\subsection*{Author contributions}
Anas H. Alzahrani conceived the URTI adaptation, directed the simulation and comparison objectives, and is responsible for the complete work. Computational implementation, analysis and manuscript preparation used the generative AI assistance described below.

\subsection*{Data and code availability}
The ancillary archive URTI\_Multiagent\_Code\_Data\_Results.zip contains the communication source, locally frozen design, compact float64 probability inputs, simulated labels and encounter identifiers, results, a representative message transcript, and verification records. It supports replay without the original model runtime. The separate frozen archive URTI\_V2\_Code\_and\_Results.zip contains the supporting predictive benchmark source, settings, summary results and checks; it omits full generated encounter files, fitted binaries, pretrained weights and runtime assets. Its settings and seeds permit regeneration in a compatible environment. Reproducing Kumo inference requires a separately prepared compatible runtime and the identified checkpoint. Accompanying README and reproducibility notes describe both routes.

\subsection*{Generative AI assistance}
OpenAI Codex assisted study planning, simulation and model code, computational execution and verification routines, interpretation, figure preparation and manuscript drafting, including the exploratory coordination extension. Numerical statements were checked against saved results and separate verification routines. The sole human author is responsible for the manuscript, references, analyses and disclosures. AI tools are not authors.

\clearpage
\begin{figure}[p]
\centering
\includegraphics[width=\linewidth,height=.74\textheight,keepaspectratio]{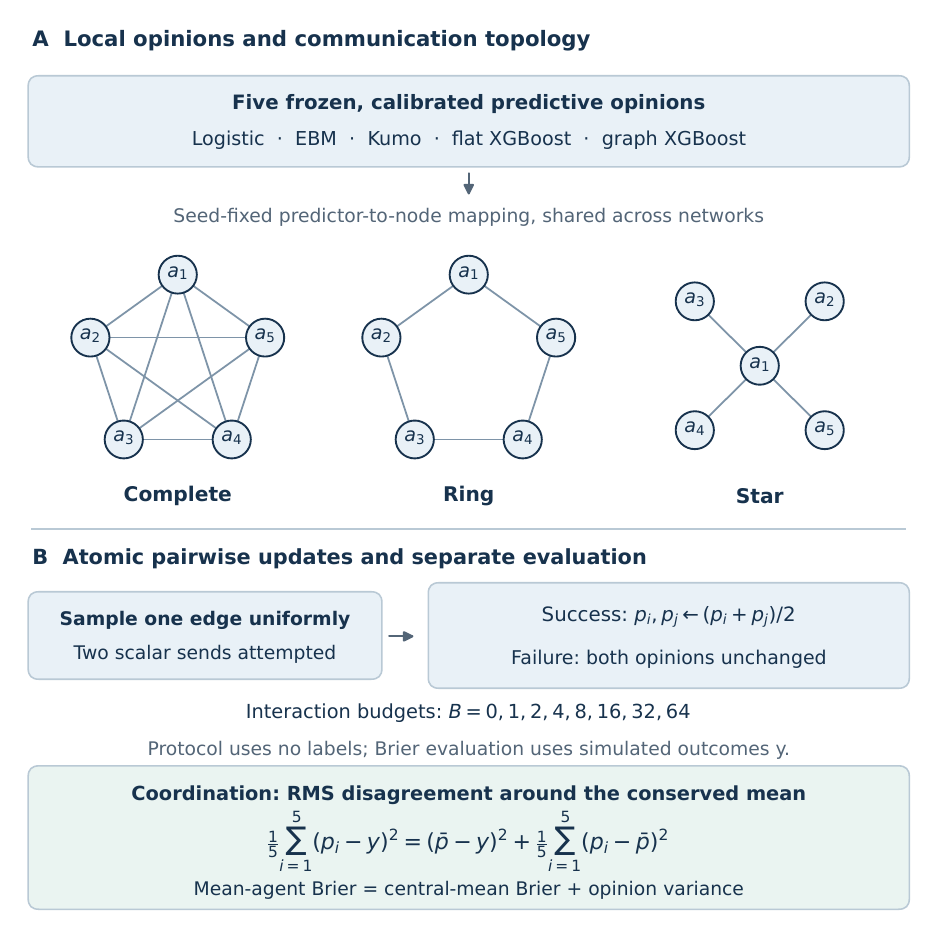}
\caption{\textbf{Five-agent communication experiment}. (A) Frozen calibrated predictions initialize five local opinions. A seed-fixed assignment maps predictors to nodes across complete, ring and star networks. (B) Each sampled edge attempts a reciprocal exchange: successful transactions replace both opinions by their previous mean, while failed transactions leave both unchanged. Communication uses no outcome labels. Evaluation measures disagreement around the conserved mean and decomposes mean-agent Brier error using simulated outcomes. Budgets count attempted pairwise interactions, each representing two scalar transmissions. The communication graphs are distinct from the feature graph inside one predictor.}\label{fig:architecture}
\end{figure}
\clearpage
\clearpage
\begin{figure}[p]
\centering
\includegraphics[width=\linewidth,height=.74\textheight,keepaspectratio]{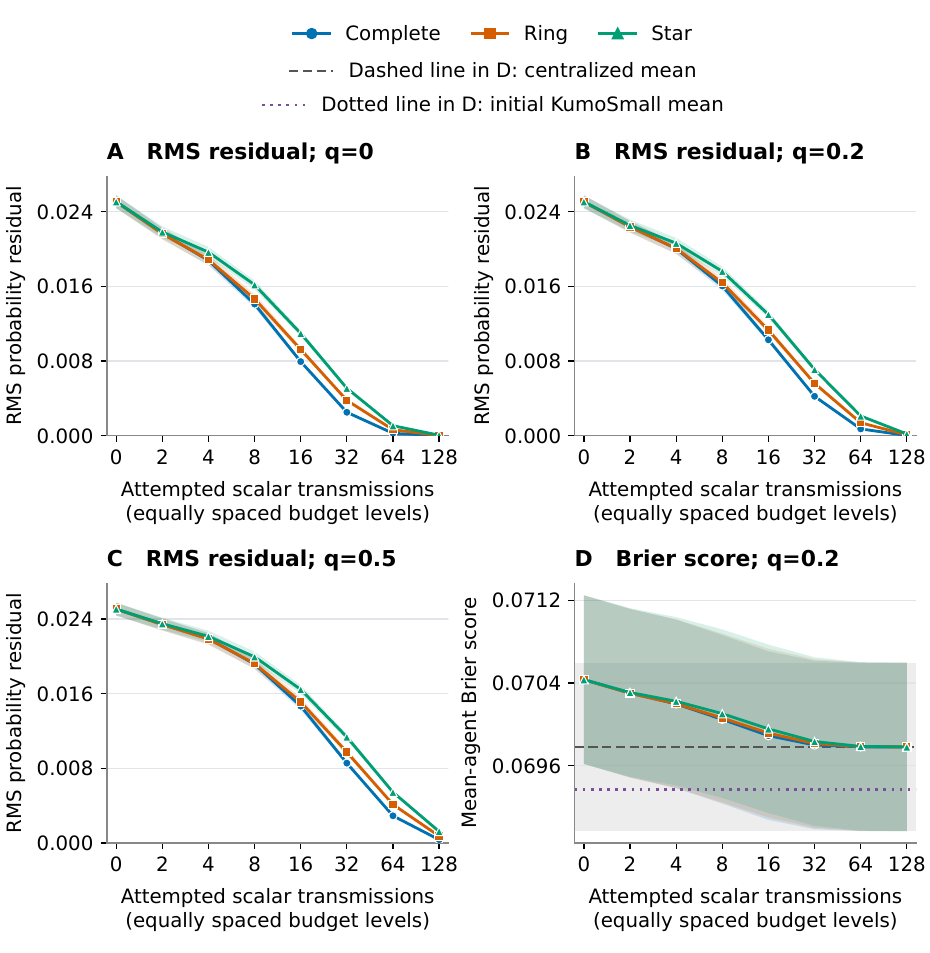}
\caption{\textbf{Finite-budget probability coordination}. Finite-budget coordination of five fixed prediction agents. Panels A--C show the mean run-specific root-mean-square (RMS) deviation of local probabilities from the centralized equal-weight mean in the base view, under whole-exchange failure probabilities q=0, 0.2, and 0.5. Panel D shows mean-agent Brier score at q=0.2; the dashed line and gray interval mark the ideal centralized equal-weight reference, whose communication cost is not modeled. The dotted line gives the initial KumoSmall mean Brier score, which is lower than the centralized pool in this experiment; its uncertainty interval is omitted for readability. Each successful exchange atomically averages the two selected endpoints; each attempt counts as two scalar transmissions. Points correspond to eight evaluated budgets, displayed at equally spaced positions; connecting lines guide the eye and are not fitted curves. Colored bands and the gray interval are pointwise conditional 95\% Monte Carlo intervals from 30 repetitions in each of three fixed generator families, weighted equally. These intervals quantify simulation uncertainty, not clinical uncertainty. Lower values indicate less disagreement or smaller mean-agent prediction error.}\label{fig:convergence}
\end{figure}
\clearpage
\clearpage
\begin{figure}[p]
\centering
\includegraphics[width=\linewidth,height=.74\textheight,keepaspectratio]{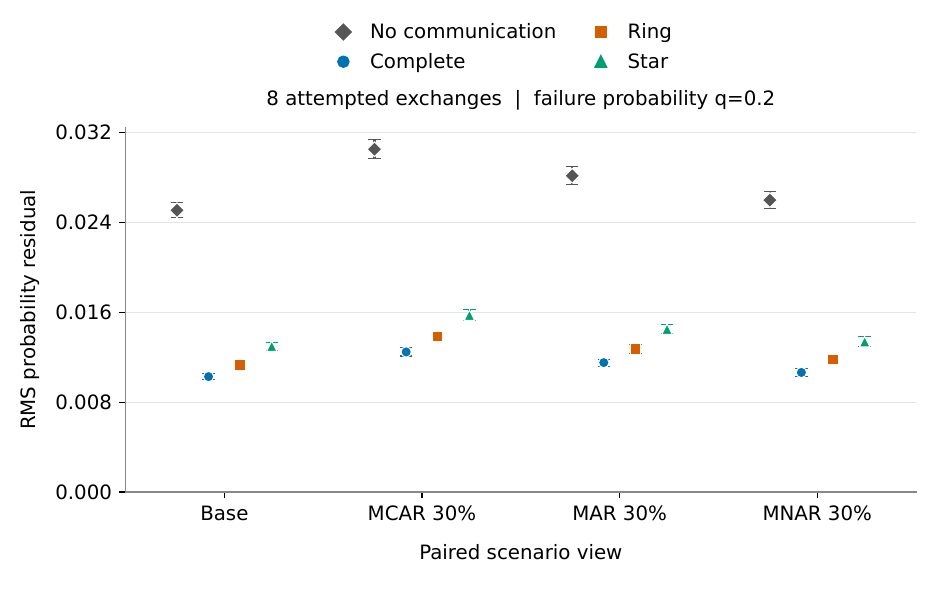}
\caption{\textbf{Coordination under missing predictors}. Coordination residual across four paired scenario views at eight attempted exchanges (16 attempted scalar transmissions) and whole-exchange failure probability q=0.2. Points show the mean run-specific RMS deviation of local probabilities from the centralized equal-weight mean; bars are pointwise conditional 95\% Monte Carlo intervals with equal weights for the three fixed generator families. The no-communication reference uses the initial agent probabilities (zero exchanges). The views reuse the same simulated encounters and are not independent patient cohorts. MCAR, MAR, and MNAR denote the three specified missingness mechanisms; 30\% applies to eligible predictor fields. Residuals are coordination measures and do not establish clinical accuracy or benefit to every individual predictor.}\label{fig:missingness}
\end{figure}
\clearpage
\clearpage\appendix\section{Supporting predictive benchmark}\label{app:benchmark}
This appendix records the completed predictive benchmark that supplied the five initial opinions. Its designated graph-versus-flat comparison preceded the exploratory communication extension. The seven-condition predictive study and separate assay experiment have a broader scope than the four-view communication experiment. All original numerical result tables are retained unchanged; table and figure numbering is updated.

\subsection{Supporting benchmark methods}
\subsubsection{Study design and scope}
We conducted a Monte Carlo methods study organized around aims, data-generating mechanisms, estimands, methods and performance measures. \cite{ref6} It represented hypothetical adults aged 18--80 years presenting with acute sore throat. No patient records, clinician participants or treatment outcomes were used. The binary primary outcome indicated whether GAS was the simulated cause of the episode; a negative outcome did not imply that all other bacterial causes were absent. The study did not cover all respiratory syndromes, antimicrobial resistance or prescribing decisions.

The completed experiment comprised 30 independently seeded runs in each of three fixed generator families. The 90 runs used 2,000 training encounters, 1,000 separate calibration encounters and 3,000 test encounters per scenario. Base prevalence was targeted at 10\%. Seven scenarios comprised the base condition, target prevalences of 5\% and 15\%, altered symptom--disease relationships, and three missingness stresses. Base and missingness-stress views shared underlying encounters; the other test sets were separately generated. This produced 1,350,000 independent simulated encounters and 2,160,000 exported encounter rows. These counts are computational sample sizes, not a patient cohort.

Following a nine-run engineering pilot, the final predictive experiment comprised 90 runs. Simulation precision was assessed using achieved Monte Carlo standard errors rather than a clinical power calculation. Local source and configuration hashes recorded the generation, fitting and analysis stages. The benchmark configuration was recorded before full model fitting, while the analysis record preceded the combined results and secondary analyses. These records document computational provenance and do not constitute external preregistration.

\subsubsection{Data-generating mechanisms}
The additive and interaction families generated correlated binary findings from independent latent severity and viral-pattern variables, then sampled the disease outcome from a specified logistic probability. The interaction family added three product terms. A separate, fixed 50,000-record reference draw per family set the intercept for each target prevalence and relationship-shift condition, so realized prevalence was approximate. In the third family, disease status and an independent presentation subtype were sampled first, followed by conditionally distributed findings; the subtype induced additional symptom dependence. All coefficients, joint distributions and subtype probabilities were hypothetical.

Twelve pre-test predictors represented age, symptom duration, fever history above 38\ensuremath{^{\circ}}C, fever during the preceding 24 hours, measured temperature, cough, coryza, purulent tonsillar exudate, tonsillar swelling, severe tonsillar inflammation, tender anterior cervical nodes and household GAS exposure. The simulator enforced consistency between current temperature above 38\ensuremath{^{\circ}}C and both fever-history fields. The outcome generator did not call the clinical-score or graph functions. Nevertheless, shared clinical intuition informed their design, so code separation does not imply complete conceptual independence. Complete-data oracle probabilities were stored for checks, excluded from every model input and not treated as achievable benchmarks for models with missing observations.

\subsubsection{Missingness and distribution shifts}
Training, calibration and base test data had a 10\% independent probability of missingness in each of nine eligible fields. Age, duration and cough were always observed, giving approximately 7.5\% missingness over all twelve feature cells. Missingness stresses used a 30\% expected rate in eligible fields, approximately 22.5\% overall. Missing completely at random (MCAR) did not depend on values. Missing at random (MAR) depended only on the three always-observed fields. Missing not at random (MNAR) depended on the value of the field that could itself be hidden, with positive findings more likely to be missing. MAR and MNAR intercepts calibrated the expected eligible-field rate conditional on complete generated covariates, without using outcome labels.

The prevalence scenarios changed the logistic intercept in the conditional families or the disease prior in the latent family. Consequently, prevalence changes did not represent an identical form of shift in every family. The presentation-shift scenario weakened the disease--finding relationship while targeting 10\% prevalence. No model or calibrator was updated using any test scenario. These are specified distribution changes, not external validation across actual clinics.

\subsubsection{Comparators and representations}
The twelve methods were a training-prevalence reference; Centor, McIsaac and FeverPAIN component scores each mapped to risk by logistic regression; multivariable regularized logistic regression; random forest; additive explainable boosting (EBM); four XGBoost representations; and Kumo Tabular Small. XGBoost representations were raw features, flat knowledge, graph summaries added to flat knowledge, and shuffled-graph summaries added to flat knowledge. \cite{ref7,ref8,ref9,ref11,ref12} Every method used the same encounters and outcome labels within a run.

Score versions were declared explicitly. Centor followed NICE component wording, including fever history above 38\ensuremath{^{\circ}}C. McIsaac used measured temperature above 38\ensuremath{^{\circ}}C as an operationalization of the temperature component displayed in the IDSA supplement, tonsillar swelling or exudate, and subtraction of one point from age 45. This measured-only fever choice is narrower than versions allowing fever history. FeverPAIN used the preceding-24-hour fever field, purulence, presentation within three days, severe tonsillar inflammation, and neither cough nor coryza. \cite{ref3,ref5} Unknown required components yielded an unknown score unless a Boolean condition could be resolved from a known input. Score-only models used the training-median score when unavailable. These were fitted synthetic GAS mappings, not published risk tables. FeverPAIN was originally derived against group A, C or G streptococci, making its GAS-only use here an adapted comparator. \cite{ref10}

The flat-knowledge representation contained 36 columns: twelve raw inputs, six unique derived primitives, fifteen primitive-availability indicators and three scores. The graph added ten unweighted means or coverage fractions across five investigator-defined finding groups. Its sixteen edges drew on clinical associations and score components, but group membership and equal weights were design choices. Means used observed findings; an entirely unobserved group retained an unknown mean. Graph and flat models therefore shared underlying information but differed in representation dimension. The graph and shuffled control both had 46 input columns; one fixed shuffle preserved source and target degrees and avoided duplicate edges. A singleton exposure summary duplicated an underlying input. No graph neural network was trained.

Logistic regression used an L2 penalty, C=1 and training-fitted standardization. Random forest used 100 trees, depth at most eight and a minimum leaf size of ten. XGBoost used 100 trees, maximum depth three, learning rate 0.07, row subsampling 0.9, minimum child weight five and L2 regularization five; column subsampling was disabled. EBM used no pairwise interactions, four outer bags and at most 500 boosting rounds, with its training-only internal validation and early-stopping defaults. \cite{ref8,ref11} Detailed settings and software versions are provided in the supplement. No test-driven hyperparameter search was performed.

Kumo used the cached Small classifier with four estimators and all 2,000 context records, following the installed structured-data-models default transformations and output normalization. Its fit operation cached in-context examples; pretrained weights were not gradient fine-tuned. Inference used float16 CUDA execution. Kumo and EBM retained native missing-value handling; other multivariable baselines used training-median imputation and missingness indicators. These choices compare complete pipelines rather than isolating an algorithm independent of preprocessing. \cite{ref9,ref12}

\subsubsection{Calibration, estimands and uncertainty}
For each fitted method, a separate sigmoid model regressed calibration-split outcomes on clipped logits of raw predictions. All test scenarios were then evaluated using both raw and calibrated probabilities. The designated primary comparison was calibrated graph-minus-flat XGBoost Brier score in the base condition; negative differences favored graph summaries. Secondary contrasts compared Kumo with logistic regression, EBM with logistic regression, and graph with shuffled-graph XGBoost. AUROC, average precision, log loss and calibration diagnostics were also recorded. Calibration-in-the-large used a fixed-slope offset intercept; joint intercept and slope were assessed separately and were not identifiable for constant predictions.

Brier differences were paired within run and averaged with equal weight across the three fixed generator families. 

For $F$ fixed families and $n_f$ runs with within-family variance $s_f^2$, the squared Monte Carlo standard error was $\sum_f(s_f^2/n_f)/F^2$. Pointwise 95\% intervals used Satterthwaite t degrees of freedom. This uncertainty describes finite simulation repetition conditional on the chosen families and reference draws, not uncertainty about clinical performance. Secondary contrasts were exploratory, without multiplicity adjustment. Threshold summaries at 0.10 and 0.20 were retained as mathematical sensitivity analyses; they were not proposed testing or prescribing cutoffs.

\subsubsection{Outcome-measurement experiment}
A separate logistic-regression experiment distinguished etiologic disease, organism presence and observed assay positivity. In a moderate scenario, assumed assay sensitivity to organism presence, specificity to absence and carriage probability among non-GAS cases were 0.85, 0.98 and 0.03. A stress scenario used 0.70, 0.95 and 0.08. These were hypothetical settings, not empirical assay-performance or Saudi carriage estimates. Verification probability depended only on recorded fever, exudate and cough, with separate verification draws for the two scenarios. Untested outcomes remained unknown.

The truth-trained logistic model was compared with models fitted and calibrated using either complete simulated assay results or only verified assay results. Verified fitting sets excluded unknown outcomes; they did not replace them with negatives. All models were evaluated on the same full base test sample against disease truth, with additional assay-target and verified-subset evaluations. The selected-label comparison changed both the number and composition of available cases, so it could not isolate a causal effect of verification bias. Complete formulae and label semantics are given in the supplement.

\subsection{Supporting benchmark results}
\subsubsection{Base-scenario performance}
All 990 conventional/\allowbreak{}EBM model fits and 90 Kumo context fits completed. The primary benchmark contained 15,120 unique method--run--scenario--variant evaluations, without exclusions. Kumo had the lowest mean calibrated base Brier score (0.069368), followed by logistic regression (0.069900) and EBM (0.070156). Their mean AUROCs were 0.830, 0.826 and 0.824, respectively. The prevalence-only reference had Brier 0.089711. Complete twelve-method results are reported in Supplementary Table S1; family-specific results and calibration diagnostics are in Supplementary Tables S5 and S6.

Graph-summary XGBoost achieved Brier 0.071003, compared with 0.071731 for flat knowledge. The primary paired difference was \ensuremath{-}0.000727, with Monte Carlo standard error 0.000071 and 95\% interval \ensuremath{-}0.000869 to \ensuremath{-}0.000586 (Supplementary Table S2). Mean graph-minus-flat differences were negative in all three generator families, ranging from \ensuremath{-}0.000382 in the additive family to \ensuremath{-}0.001173 in the latent-mixture family (Supplementary Figure S1). The graph-minus-shuffled difference was \ensuremath{-}0.000430 (\ensuremath{-}0.000556 to \ensuremath{-}0.000303). The graph model nevertheless had higher average error than Kumo, logistic regression and EBM.

The secondary Kumo-minus-logistic difference was \ensuremath{-}0.000533 (\ensuremath{-}0.000643 to \ensuremath{-}0.000422); EBM-minus-logistic was +0.000256 (+0.000120 to +0.000391). These were small absolute probability-error differences. The score-based synthetic mappings had higher base errors: 0.078492 for Centor, 0.081349 for McIsaac and 0.081939 for FeverPAIN. They were given fewer predictive inputs, used declared score operationalizations and faced the chosen missingness process; these results are not evidence that the clinical scores fail in their intended populations.

\subsubsection{Robustness and outcome measurement}
Among the five principal methods displayed in Supplementary Table S3, Kumo had the lowest mean calibrated Brier in each specified stress condition. Under 30\% eligible-field MNAR missingness, mean Brier was 0.075326 for Kumo, 0.077258 for logistic regression, 0.078267 for EBM, 0.076990 for graph-summary XGBoost and 0.080793 for flat-knowledge XGBoost. The paired missingness views are summarized in Supplementary Figure S2. Under the changed presentation relationship, errors were much closer: 0.088879, 0.089069, 0.089023, 0.089342 and 0.089346, respectively. Absolute Brier scores across prevalence settings should not be interpreted as differences in model quality because outcome frequency also changes task difficulty.

The outcome-measurement experiment added 360 logistic fits and 2,160 evaluations. Against disease truth on the full base test set, training on complete moderate-scenario assay labels increased mean Brier from 0.069900 to 0.073623; using only verified assay labels increased it to 0.076257. Corresponding stress-scenario values were 0.081703 and 0.084656 (Supplementary Table S4). Verified models used approximately 651 or 648 training cases, versus 2,000 for full-label models. Mean assay positivity was 12.5\% or 16.1\% over all test cases and 15.2\% or 18.2\% among verified cases, while disease prevalence was approximately 10\%. These patterns illustrate target mismatch, selection and reduced sample size under the specified mechanisms.

\subsubsection{Computational execution and numerical verification}
The conventional/\allowbreak{}EBM benchmark took 191.1 seconds, and the complete Kumo container invocation produced 1.98 million predictions, including calibration cases, in 480.1 seconds. Kumo ran locally in a network-disabled container on an NVIDIA GB10 GPU; maximum PyTorch-allocated memory was 250.33 MiB, excluding other runtime allocations. A warmed, transfer-inclusive XGBoost probe at 2,000 training rows measured median CPU and CUDA times of 0.01275 and 0.04927 seconds, respectively, supporting CPU XGBoost for this workload. Timings reflect a shared workstation, not a dedicated hardware comparison.

Twenty-three automated tests passed. Separate verification routines recomputed all 15,120 primary and 2,160 secondary Brier scores, with maximum absolute discrepancies of 4.16\ensuremath{\times}10\textsuperscript{-}\textsuperscript{1}\textsuperscript{7} and 5.55\ensuremath{\times}10\textsuperscript{-}\textsuperscript{1}\textsuperscript{7}. All 990 baseline artifacts were present; prediction replay checks covered 33 representative baseline artifacts spanning every method and family, plus fifteen secondary-model artifacts. Kumo source/\allowbreak{}output alignment and sigmoid recalculation were checked, but full Kumo inference was not independently repeated. Raw XGBoost CSV probabilities were restored to their original float32 type for exact arithmetic checks. These procedures verify internal numerical consistency, not scientific realism.

\section{Simulation mechanisms and implementation settings}
\subsection*{Simulation mechanisms and parameter ledger}
Notation: \ensuremath{\sigma}(z)=1/\allowbreak{}(1+exp(\ensuremath{-}z)). A is age, D symptom duration, F fever history >38\ensuremath{^{\circ}}C, R fever within 24 hours, T current temperature, C cough, V coryza, E purulent exudate, W swelling, I severe inflammation, N tender anterior cervical nodes and H household contact. Age is discrete uniform 18--80 and duration discrete uniform 1--10 days. All distributional values below are design assumptions.

\subsubsection*{Conditional additive and interaction families}
Let latent severity S and viral-pattern U be independent standard normal variables. Each primary binary observation is drawn with probability \ensuremath{\sigma}(a+bS+cU). Coefficients (a,b,c) are: F (\ensuremath{-}0.6,0.9,0); C (\ensuremath{-}0.1,0.1,1.0); V (\ensuremath{-}0.5,0,1.1); E (\ensuremath{-}1.4,0.85,\ensuremath{-}0.1); I (\ensuremath{-}0.5,0.9,0); N (\ensuremath{-}1.0,0.7,0). H is Bernoulli(0.12). Binary draws are conditionally independent given S and U.

The additive predictor is
\begin{align*}
z={}&0.75F+0.95E+0.65N-0.8C-0.65V+0.75H\\
 &-0.012(A-40)+0.25I-0.04(D-3)+0.15W+0.12R.
\end{align*}
The interaction predictor adds $0.9EN+0.65F(1-C)-0.8CV$.
Disease $Y$ is Bernoulli$(\sigma(\alpha+z))$.
For each family, target prevalence and relationship-shift condition, \ensuremath{\alpha} is solved by bisection using a fixed independent 50,000-record covariate reference sample (seed 107 for additive and 108 for interactions). The presentation shift replaces z with 0.55z\ensuremath{-}0.7E+0.35C and resolves \ensuremath{\alpha} for target prevalence 0.10. Prevalence targets describe the reference expectation; sample prevalence varies.

\subsubsection*{Latent disease/\allowbreak{}subtype family}
Y is Bernoulli(p) and subtype B is independently Bernoulli(0.30). For findings (F,C,V,E,I,N,H), base probabilities when Y=0 are (0.25,0.65,0.55,0.15,0.25,0.20,0.08), and when Y=1 they are (0.65,0.20,0.15,0.55,0.60,0.50,0.25). Add (B\ensuremath{-}0.30) times (0.14,0.08,0.08,0.12,0.10,0.10,0), then clip each probability to [0.02,0.98]. Findings are conditionally independent given Y and B. Presentation shift changes the Y=1 base rates to 0.55 times their original values plus 0.45 times the Y=0 rates. The oracle is calculated by Bayes' rule after marginalizing B and remains outside model inputs.

\subsubsection*{Derived observations in all families}
T is sampled from Normal(36.8+1.45F,0.38\textsuperscript{2}), clipped to [35.5,40.5]\ensuremath{^{\circ}}C, and capped at 38\ensuremath{^{\circ}}C when F=0. An initial R is Bernoulli(0.85) if F=1 and Bernoulli(0.10) otherwise; R is then set to 1 whenever current T>38\ensuremath{^{\circ}}C. W is Bernoulli(0.12+0.40E+0.28I). In the latent family, these derived variables add no disease information conditional on the primary findings and age/\allowbreak{}duration. This is a specified simplification, not a claim about real fever or swelling measurement.

\subsection*{Observation and outcome-measurement processes}
Eligible masked fields are all predictors except A, D and C. MCAR masks each eligible cell independently with probability 0.10 in development/\allowbreak{}base data or 0.30 in the MCAR stress. For MAR, each eligible field uses missingness probability \ensuremath{\sigma}(\ensuremath{\alpha}\_M+0.035(A\ensuremath{-}45)+0.9C+0.12(D\ensuremath{-}3)). For MNAR, it uses \ensuremath{\sigma}(\ensuremath{\alpha}\_N+1.8X), where X is that field's complete binary value, or 1(T>37.5) for temperature. Each intercept is solved so the expected mask fraction is 0.30 across that generated complete field. Separate random draws determine masks. The base and three masked stress views retain identical underlying labels and case IDs, but are not necessarily nested masks.

For assay scenario j, carriage occurs only when Y=0, with probability c\_j. Organism presence O is 1 if Y=1 or carriage is present. Full assay result Z is Bernoulli(Se\_j) when O=1 and Bernoulli(1\ensuremath{-}Sp\_j) otherwise. Moderate parameters (Se,Sp,c) are (0.85,0.98,0.03); stress parameters are (0.70,0.95,0.08). Neither setting was fitted from clinical test data. These Se/\allowbreak{}Sp values relate to organism presence, not etiologic disease.

Let F* and E* equal recorded fever history and exudate, with 0.5 used only in the verification formula if missing. Verification probability is q=0.12+0.65\ensuremath{\sigma}(F*+E*\ensuremath{-}0.9C\ensuremath{-}1). Verification is sampled independently within each assay world; the observed label equals Z if verified and is unknown otherwise. Only observed labels enter verified-model fitting and calibration. The model input list never includes carriage, O, Z, verification status, the complete-data oracle, identifiers or hidden outcomes. Primary fitting continues to use the simulator's Y in a separate target column.

\subsection*{Feature representations and model settings}
The flat representation includes raw variables; six derived primitives (no cough, attendance within three days, neither cough nor coryza, swelling or exudate, current temperature >38\ensuremath{^{\circ}}C and age \ensuremath{\geq}45); fifteen primitive-availability indicators; and three component scores. Graph groups are GAS pattern, viral pattern, inflammation pattern, exposure context and FeverPAIN context. Sixteen source--group edges are stored with source links in the reproducibility package. Equal weights encode no likelihood ratios or causal strengths. A degree-preserving target shuffle with seed 20261107 rejects duplicate edges and identical assignments; only that one shuffled graph is evaluated.

Raw, flat and graph model matrices contain 12, 36 and 46 input columns, respectively, before model-specific imputation/\allowbreak{}indicator expansion. Both graph variants have the same dimension. SimpleImputer uses training medians, keeps all-empty features, and adds indicators for non-score tabular models. StandardScaler is fitted only for logistic models. EBM and Kumo retain native missing values. Clinical-score models have a single median-filled score input. Clinical-score probabilities are learned on synthetic data and are then separately sigmoid-calibrated.

Logistic models use L2-regularized scikit-learn LogisticRegression(C=1, solver=lbfgs, max\_iter=500). The sigmoid calibrator uses C=10\textsuperscript{6}, solver=lbfgs and max\_iter=500; prediction logits are clipped using probability bounds [10\textsuperscript{-}\textsuperscript{6},1\ensuremath{-}10\textsuperscript{-}\textsuperscript{6}]. Random forest uses 100 trees, max\_depth=8, min\_samples\_leaf=10, two jobs, bootstrap=True, criterion=gini and max\_features=sqrt. Seed-dependent settings are recorded per run. The prevalence-only reference is constant; its calibrated output is adjusted using the separate calibration split.

XGBoost parameters are n\_estimators=100, max\_depth=3, learning\_rate=0.07, subsample=0.9, colsample\_bytree=1.0, min\_child\_weight=5, reg\_lambda=5, objective=binary:logistic, eval\_metric=logloss, tree\_method=hist, device=cpu and n\_jobs=2. CPU was selected after a workload-specific CUDA comparison. The measured CPU/\allowbreak{}CUDA probe includes fitting and prediction on training records and is not an accuracy-validation experiment.

EBM parameters are interactions=0, outer\_bags=4, max\_rounds=500, n\_jobs=2 and missing=separate. Saved-model defaults include validation\_size=0.15, learning\_rate=0.015, early\_stopping\_rounds=100, max\_bins=1024, max\_leaves=2, min\_samples\_leaf=4 and smoothing\_rounds=75. Its internal validation records are drawn from the 2,000 training encounters, not the separate 1,000-row calibration split. Missing-bin contributions are retained in term\_scores even though standard EBM visualizations omit those bins.

Kumo uses structured-data-models 0.2.1, the cached nvidia/\allowbreak{}Kumo-Tabular Small classification checkpoint at revision v1.0.1, four estimators, estimator\_batch\_size=1 and prediction batches of 128. Predictors are represented as numerical TableTensor columns; target classes are categorical strings, and probability of named class ``1'' is explicitly extracted. Both class probabilities are checked for finite values, bounds and unit sum. The installed default recipe applies training-context transformations and ensemble output normalization. In-context fitting does not gradient fine-tune the checkpoint. Model revision, checkpoint commit and SHA-256 are recorded in the original inference manifest. The current online model card is background documentation, not evidence that a newer release was run.

\subsection*{Estimands, calibration and numerical checks}
For each run, method and scenario, Brier equals the mean of $(Y-p)^2$.
Primary contrasts use calibrated predictions and pair methods on the same encounters.
Each of $F=3$ fixed families contributes equally; within-family means use $n_f=30$ independent repetitions.
Let $v_f=s_f^2/(n_f F^2)$. Then
\[
\operatorname{MCSE}=\sqrt{\sum_f v_f},\qquad
\nu=\frac{(\sum_f v_f)^2}{\sum_f[v_f^2/(n_f-1)]}.
\]
Intervals use the corresponding 97.5th $t$ quantile with Satterthwaite degrees of freedom $\nu$.
This treats reference-draw mechanisms as fixed. It does not average over arbitrary possible clinical worlds or patients.

Calibration-in-the-large solves mean[\ensuremath{\sigma}(logit(p)+a)]=mean(Y) with slope fixed to one. Joint calibration intercept and slope are fitted as a diagnostic regression on test labels; these diagnostics do not alter saved test predictions. Joint parameters are not identifiable for the constant model. Average precision, AUROC and log loss are reported in the full results. Available confusion-matrix and mathematical net-benefit summaries at thresholds 0.10/\allowbreak{}0.20 do not establish a clinical utility function or actionable threshold.

Twenty-three automated tests cover score boundaries/\allowbreak{}unknowns, feature leakage, generator reproducibility/\allowbreak{}consistency, missingness, assay-label exclusion, model serialization, metric arithmetic and Monte Carlo summaries. Separate routines checked all primary and secondary Brier evaluations against saved per-case predictions. The primary check verified 810 encounter-file hashes and 2,160 Kumo/\allowbreak{}calibration provenance hashes. All 990 conventional/\allowbreak{}EBM artifacts were present; 33 representative artifacts and fifteen secondary artifacts were replayed. Kumo raw inference was not rerun independently; saved class alignment, hashes and sigmoid probabilities were checked. A verified-resume check skipped three completed Kumo runs without refitting.

An initial, unanalysed generator version allowed measured fever above 38\ensuremath{^{\circ}}C without fever in the preceding 24 hours. The generator was corrected and all active data were regenerated before fitting. The unused version and original hash record were preserved separately. A later configuration-guard amendment did not change numerical generation. Raw XGBoost CSV values use float32 representations; the verifier reconstructs this type before exact comparisons. Other evaluated probability columns use float64. These are implementation and provenance checks, not clinical validation.

\subsection*{Design development and analysis transparency}
A nine-run engineering pilot preceded the completed experiment. The final design changed predictor semantics, development-set missingness, graph controls, repetition count and comparators. Pilot results were not pooled with the reported benchmark, and differences between the pilot and final experiment cannot be attributed to a single modification. Local source and configuration records document the stages of development and analysis; the study was not externally preregistered.

The reported scope comprises 30 repetitions per family, 2,000 training, 1,000 calibration and 3,000 test encounters per scenario, fixed 10\% development prevalence, and separate prevalence, presentation, missingness and outcome-measurement experiments. Learning curves, multiple graph randomizations, a trained graph neural network, clinician assessment, policy outcomes and explanation-fidelity experiments were not performed.

\subsection*{Environment and access}
The numerical environment used Python 3.12.3, NumPy 2.5.4, pandas 3.1.0, SciPy 1.18.1, scikit-learn 1.9.1, XGBoost 3.4.1, interpret-core 0.7.8 and joblib 1.6.0. Exact local package and checkpoint provenance is retained in the study. The project-owned environment referenced an existing numerical package directory without modifying it. Kumo used cached weights in a network-disabled, resource-bounded container. Category-name sorting used a CPU fallback where cuDF was absent; prediction remained CUDA.

The accompanying ancillary archive URTI\_V2\_Code\_and\_Results.zip contains simulation source, configuration files, summary results and verification records. It omits full generated encounter files, fitted binaries, pretrained weights and runtime assets. The settings and seeds permit regeneration in a compatible environment. The archive is not a portable trained clinical system. Reproducing Kumo inference requires a separately prepared compatible runtime and the identified pretrained checkpoint. The ancillary file REPRODUCIBILITY.txt describes the required environment and path adaptations.

The later communication extension, including its distinct local freeze and complete fixed inputs, is supplied in URTI\_Multiagent\_Code\_Data\_Results.zip. Its configuration was recorded after inspection of the benchmark results; it is exploratory. The original model benchmark archive remains byte-identical to its frozen version.

\clearpage\section{Supporting tables and figures}\renewcommand{\thetable}{S\arabic{table}}\renewcommand{\theHtable}{S\arabic{table}}\setcounter{table}{0}

\begingroup\small
\begin{longtable}{@{}>{\raggedright\arraybackslash}p{0.4200\linewidth}>{\raggedright\arraybackslash}p{0.1200\linewidth}>{\raggedright\arraybackslash}p{0.1200\linewidth}>{\raggedright\arraybackslash}p{0.1200\linewidth}>{\raggedright\arraybackslash}p{0.1200\linewidth}@{}}
\caption{Base-scenario performance of the 12 calibrated methods}\label{tab:original_T1}\\
\toprule
\textbf{Method} & \textbf{Brier} & \textbf{MCSE} & \textbf{AUROC} & \textbf{AP} \\
\midrule\endfirsthead
\caption[]{Base-scenario performance of the 12 calibrated methods (continued)}\\
\toprule
\textbf{Method} & \textbf{Brier} & \textbf{MCSE} & \textbf{AUROC} & \textbf{AP} \\
\midrule\endhead
\midrule\multicolumn{5}{r}{Continued on next page}\\\endfoot
\bottomrule\endlastfoot
Kumo Tabular Small & 0.06937 & 0.000411 & 0.82972 & 0.44962 \\
Logistic regression & 0.06990 & 0.000413 & 0.82648 & 0.44213 \\
Explainable boosting (EBM) & 0.07016 & 0.000413 & 0.82439 & 0.43831 \\
XGBoost: graph & 0.07100 & 0.000415 & 0.81871 & 0.42502 \\
XGBoost: shuffled graph & 0.07143 & 0.000408 & 0.81639 & 0.41868 \\
XGBoost: raw & 0.07148 & 0.000417 & 0.81627 & 0.42083 \\
XGBoost: flat knowledge & 0.07173 & 0.000415 & 0.81509 & 0.41578 \\
Random forest & 0.07198 & 0.000421 & 0.81446 & 0.41524 \\
Centor + logistic & 0.07849 & 0.000434 & 0.72656 & 0.27339 \\
McIsaac + logistic & 0.08135 & 0.000447 & 0.71358 & 0.23315 \\
FeverPAIN + logistic & 0.08194 & 0.000468 & 0.69784 & 0.22856 \\
Training prevalence & 0.08971 & 0.000533 & 0.50000 & 0.09950 \\
\end{longtable}
\noindent Equal-weight mean over three fixed hypothetical generator families, each with 30 independent replicates (90 runs). All methods used a separate sigmoid calibration sample. Lower Brier is better; higher AUROC and average precision (AP) are better. MCSE is the Monte Carlo standard error of mean Brier, not uncertainty about clinical performance. Zero primary fit failures; no runs omitted. Score comparators are logistic mappings fitted to the specified clinical score, not published score-to-risk tables. The constant model has AUROC 0.5.
\par\endgroup
\clearpage

\begingroup\small
\begin{longtable}{@{}>{\raggedright\arraybackslash}p{0.3200\linewidth}>{\raggedright\arraybackslash}p{0.1500\linewidth}>{\raggedright\arraybackslash}p{0.1200\linewidth}>{\raggedright\arraybackslash}p{0.3100\linewidth}@{}}
\caption{Paired differences in base-scenario calibrated Brier score}\label{tab:original_T2}\\
\toprule
\textbf{Contrast} & \textbf{Mean difference} & \textbf{MCSE} & \textbf{95\% MC interval} \\
\midrule\endfirsthead
\caption[]{Paired differences in base-scenario calibrated Brier score (continued)}\\
\toprule
\textbf{Contrast} & \textbf{Mean difference} & \textbf{MCSE} & \textbf{95\% MC interval} \\
\midrule\endhead
\midrule\multicolumn{4}{r}{Continued on next page}\\\endfoot
\bottomrule\endlastfoot
Graph \ensuremath{-} flat XGBoost & -0.000727 & 0.000071 & [-0.000869, -0.000586] \\
Kumo \ensuremath{-} logistic & -0.000533 & 0.000055 & [-0.000643, -0.000422] \\
EBM \ensuremath{-} logistic & 0.000256 & 0.000068 & [0.000120, 0.000391] \\
Graph \ensuremath{-} shuffled XGBoost & -0.000430 & 0.000063 & [-0.000556, -0.000303] \\
\end{longtable}
\noindent Negative values favor the first method. Differences were calculated within the same run before aggregation, with equal weights for the three fixed generator families. Intervals are pointwise Monte Carlo t intervals using Satterthwaite degrees of freedom; they are conditional on the specified simulation worlds and are not multiplicity-adjusted or clinical confidence intervals. Graph minus flat XGBoost was the primary contrast; the other three contrasts are exploratory. All 90 paired runs are included.
\par\endgroup
\clearpage

\begingroup\small
\begin{longtable}{@{}>{\raggedright\arraybackslash}p{0.2500\linewidth}>{\raggedright\arraybackslash}p{0.1300\linewidth}>{\raggedright\arraybackslash}p{0.1300\linewidth}>{\raggedright\arraybackslash}p{0.1300\linewidth}>{\raggedright\arraybackslash}p{0.1300\linewidth}>{\raggedright\arraybackslash}p{0.1300\linewidth}@{}}
\caption{Calibrated mean Brier score across all seven test conditions}\label{tab:original_T3}\\
\toprule
\textbf{Test condition} & \textbf{Logistic} & \textbf{Kumo} & \textbf{EBM} & \textbf{Flat XGB} & \textbf{Graph XGB} \\
\midrule\endfirsthead
\caption[]{Calibrated mean Brier score across all seven test conditions (continued)}\\
\toprule
\textbf{Test condition} & \textbf{Logistic} & \textbf{Kumo} & \textbf{EBM} & \textbf{Flat XGB} & \textbf{Graph XGB} \\
\midrule\endhead
\midrule\multicolumn{6}{r}{Continued on next page}\\\endfoot
\bottomrule\endlastfoot
Base & 0.06990 & 0.06937 & 0.07016 & 0.07173 & 0.07100 \\
Prevalence 5\% & 0.04273 & 0.04249 & 0.04280 & 0.04351 & 0.04332 \\
Prevalence 15\% & 0.09862 & 0.09799 & 0.09905 & 0.10102 & 0.10028 \\
Presentation shift & 0.08907 & 0.08888 & 0.08902 & 0.08935 & 0.08934 \\
MCAR 30\% & 0.07370 & 0.07195 & 0.07357 & 0.07597 & 0.07410 \\
MAR 30\% & 0.07230 & 0.07111 & 0.07230 & 0.07424 & 0.07294 \\
MNAR 30\% & 0.07726 & 0.07533 & 0.07827 & 0.08079 & 0.07699 \\
\end{longtable}
\noindent Each cell is an equal-family mean over 90 runs; lower is better. Prevalence targets are simulation assumptions. Training, calibration and base tests have 10\% MCAR missingness per eligible field. The three missingness stresses target 30\% per eligible field (nine of twelve fields; about 22.5\% of all feature cells), with observed age, duration and cough as anchors; realized rates vary. MCAR, MAR and MNAR are masked views of the same base encounters and outcomes and must not be treated as independent patient samples. EBM, explainable boosting machine; XGB, XGBoost. Scenario comparisons are descriptive.
\par\endgroup
\clearpage

\begingroup\small
\begin{longtable}{@{}>{\raggedright\arraybackslash}p{0.1000\linewidth}>{\raggedright\arraybackslash}p{0.3000\linewidth}>{\raggedright\arraybackslash}p{0.1300\linewidth}>{\raggedright\arraybackslash}p{0.1300\linewidth}>{\raggedright\arraybackslash}p{0.1200\linewidth}>{\raggedright\arraybackslash}p{0.1200\linewidth}@{}}
\caption{Secondary logistic-regression audit of labels and selective testing}\label{tab:original_T4}\\
\toprule
\textbf{Assay world} & \textbf{Training labels /\allowbreak{} sample} & \textbf{Mean train N} & \textbf{Mean cal. N} & \textbf{Brier} & \textbf{MCSE} \\
\midrule\endfirsthead
\caption[]{Secondary logistic-regression audit of labels and selective testing (continued)}\\
\toprule
\textbf{Assay world} & \textbf{Training labels /\allowbreak{} sample} & \textbf{Mean train N} & \textbf{Mean cal. N} & \textbf{Brier} & \textbf{MCSE} \\
\midrule\endhead
\midrule\multicolumn{6}{r}{Continued on next page}\\\endfoot
\bottomrule\endlastfoot
Moderate & True outcome; full sample & 2000.0 & 1000.0 & 0.06990 & 0.000413 \\
Moderate & Test label; full sample & 2000.0 & 1000.0 & 0.07362 & 0.000412 \\
Moderate & Test label; selected sample & 651.1 & 325.3 & 0.07626 & 0.000469 \\
Stress & True outcome; full sample & 2000.0 & 1000.0 & 0.06990 & 0.000413 \\
Stress & Test label; full sample & 2000.0 & 1000.0 & 0.08170 & 0.000467 \\
Stress & Test label; selected sample & 648.4 & 323.2 & 0.08466 & 0.000544 \\
\end{longtable}
\noindent All predictions are evaluated against latent etiologic truth in all 3,000 base test encounters per run, after separate calibration using each regime's labels. Equal-family means over 90 runs. Hypothetical moderate assay: sensitivity 0.85, specificity 0.98, carriage 0.03; stress assay: 0.70, 0.95, 0.08. Carriage applies among etiologically negative cases. Truth-full predictions are shared across assay worlds, so the repeated reference rows are not independent. Selected-sample models change both case mix and sample size; this design does not isolate a pure selection-bias effect. No real assay-performance claim is made.
\par\endgroup
\clearpage

\begingroup\small
\begin{longtable}{@{}>{\raggedright\arraybackslash}p{0.1800\linewidth}>{\raggedright\arraybackslash}p{0.3600\linewidth}>{\raggedright\arraybackslash}p{0.1200\linewidth}>{\raggedright\arraybackslash}p{0.1200\linewidth}>{\raggedright\arraybackslash}p{0.1200\linewidth}@{}}
\caption{Base-scenario calibrated performance by generator family}\label{tab:original_S1}\\
\toprule
\textbf{Generator family} & \textbf{Method} & \textbf{Brier} & \textbf{MCSE} & \textbf{AUROC} \\
\midrule\endfirsthead
\caption[]{Base-scenario calibrated performance by generator family (continued)}\\
\toprule
\textbf{Generator family} & \textbf{Method} & \textbf{Brier} & \textbf{MCSE} & \textbf{AUROC} \\
\midrule\endhead
\midrule\multicolumn{5}{r}{Continued on next page}\\\endfoot
\bottomrule\endlastfoot
Additive & Kumo Tabular Small & 0.08067 & 0.000839 & 0.74612 \\
Additive & Logistic regression & 0.08090 & 0.000846 & 0.74265 \\
Additive & Explainable boosting (EBM) & 0.08113 & 0.000832 & 0.74000 \\
Additive & XGBoost: graph & 0.08189 & 0.000854 & 0.73124 \\
Additive & XGBoost: shuffled graph & 0.08213 & 0.000834 & 0.72845 \\
Additive & XGBoost: raw & 0.08218 & 0.000850 & 0.72714 \\
Additive & XGBoost: flat knowledge & 0.08227 & 0.000830 & 0.72707 \\
Additive & Random forest & 0.08214 & 0.000858 & 0.72937 \\
Additive & Centor + logistic & 0.08417 & 0.000837 & 0.66641 \\
Additive & McIsaac + logistic & 0.08480 & 0.000855 & 0.66316 \\
Additive & FeverPAIN + logistic & 0.08542 & 0.000871 & 0.65346 \\
Additive & Training prevalence & 0.08865 & 0.000915 & 0.50000 \\
Interactions & Kumo Tabular Small & 0.07477 & 0.000583 & 0.82144 \\
Interactions & Logistic regression & 0.07536 & 0.000574 & 0.81742 \\
Interactions & Explainable boosting (EBM) & 0.07553 & 0.000610 & 0.81519 \\
Interactions & XGBoost: graph & 0.07604 & 0.000598 & 0.81256 \\
Interactions & XGBoost: shuffled graph & 0.07640 & 0.000607 & 0.81028 \\
Interactions & XGBoost: raw & 0.07624 & 0.000601 & 0.81045 \\
Interactions & XGBoost: flat knowledge & 0.07667 & 0.000621 & 0.80801 \\
Interactions & Random forest & 0.07638 & 0.000592 & 0.80852 \\
Interactions & Centor + logistic & 0.08039 & 0.000535 & 0.72458 \\
Interactions & McIsaac + logistic & 0.08293 & 0.000596 & 0.71720 \\
Interactions & FeverPAIN + logistic & 0.08538 & 0.000639 & 0.69321 \\
Interactions & Training prevalence & 0.09143 & 0.000660 & 0.50000 \\
Latent mixture & Kumo Tabular Small & 0.05266 & 0.000689 & 0.92162 \\
Latent mixture & Logistic regression & 0.05344 & 0.000698 & 0.91938 \\
Latent mixture & Explainable boosting (EBM) & 0.05381 & 0.000686 & 0.91798 \\
Latent mixture & XGBoost: graph & 0.05508 & 0.000682 & 0.91233 \\
Latent mixture & XGBoost: shuffled graph & 0.05577 & 0.000657 & 0.91043 \\
Latent mixture & XGBoost: raw & 0.05602 & 0.000696 & 0.91121 \\
Latent mixture & XGBoost: flat knowledge & 0.05626 & 0.000688 & 0.91018 \\
Latent mixture & Random forest & 0.05743 & 0.000713 & 0.90548 \\
Latent mixture & Centor + logistic & 0.07092 & 0.000842 & 0.78870 \\
Latent mixture & McIsaac + logistic & 0.07631 & 0.000844 & 0.76038 \\
Latent mixture & FeverPAIN + logistic & 0.07501 & 0.000898 & 0.74684 \\
Latent mixture & Training prevalence & 0.08905 & 0.001131 & 0.50000 \\
\end{longtable}
\noindent Thirty independent replicate runs per family and method, without omitted runs. MCSE describes simulation sampling error of mean Brier within the specified family. Family order and aggregate ranks are descriptive; hypothetical worlds are not sampled clinical populations.
\par\endgroup
\clearpage

\begingroup\small
\begin{longtable}{@{}>{\raggedright\arraybackslash}p{0.3300\linewidth}>{\raggedright\arraybackslash}p{0.1300\linewidth}>{\raggedright\arraybackslash}p{0.1500\linewidth}>{\raggedright\arraybackslash}p{0.1500\linewidth}>{\raggedright\arraybackslash}p{0.1400\linewidth}@{}}
\caption{Base-scenario calibration diagnostics after sigmoid calibration}\label{tab:original_S2}\\
\toprule
\textbf{Method} & \textbf{CITL} & \textbf{Joint intercept} & \textbf{Joint slope} & \textbf{Identified /\allowbreak{} total} \\
\midrule\endfirsthead
\caption[]{Base-scenario calibration diagnostics after sigmoid calibration (continued)}\\
\toprule
\textbf{Method} & \textbf{CITL} & \textbf{Joint intercept} & \textbf{Joint slope} & \textbf{Identified /\allowbreak{} total} \\
\midrule\endhead
\midrule\multicolumn{5}{r}{Continued on next page}\\\endfoot
\bottomrule\endlastfoot
Kumo Tabular Small & 0.031 & 0.066 & 1.017 & 90/\allowbreak{}90 \\
Logistic regression & 0.029 & 0.053 & 1.010 & 90/\allowbreak{}90 \\
Explainable boosting (EBM) & 0.030 & 0.057 & 1.012 & 90/\allowbreak{}90 \\
XGBoost: graph & 0.031 & 0.069 & 1.018 & 90/\allowbreak{}90 \\
XGBoost: shuffled graph & 0.034 & 0.070 & 1.016 & 90/\allowbreak{}90 \\
XGBoost: raw & 0.033 & 0.058 & 1.010 & 90/\allowbreak{}90 \\
XGBoost: flat knowledge & 0.034 & 0.068 & 1.015 & 90/\allowbreak{}90 \\
Random forest & 0.032 & 0.070 & 1.016 & 90/\allowbreak{}90 \\
Centor + logistic & 0.026 & 0.108 & 1.040 & 90/\allowbreak{}90 \\
McIsaac + logistic & 0.028 & 0.090 & 1.030 & 90/\allowbreak{}90 \\
FeverPAIN + logistic & 0.025 & 0.075 & 1.023 & 90/\allowbreak{}90 \\
Training prevalence & 0.022 & NA & NA & 0/\allowbreak{}90 \\
\end{longtable}
\noindent CITL is the mean calibration-in-the-large intercept fitted with the logit prediction as an offset (ideal 0). Joint intercept and slope are estimated together (ideals 0 and 1). These distinct intercepts are not interchangeable. Values are equal-family means of test-sample diagnostic estimates, not pooled fits. The constant training-prevalence model has 90/\allowbreak{}90 nonidentifiable joint fits because prediction logits are constant; its joint values are NA, not zero, while CITL remains identifiable. No unexpected diagnostic-fit failures occurred and no failed or undefined slope was silently dropped.
\par\endgroup
\clearpage

\begingroup\small
\begin{longtable}{@{}>{\raggedright\arraybackslash}p{0.3300\linewidth}>{\raggedright\arraybackslash}p{0.5700\linewidth}@{}}
\caption{Predictor definitions}\label{tab:dictionary}\\
\toprule
\textbf{Field} & \textbf{Meaning} \\
\midrule\endfirsthead
\caption[]{Predictor definitions (continued)}\\
\toprule
\textbf{Field} & \textbf{Meaning} \\
\midrule\endhead
\midrule\multicolumn{2}{r}{Continued on next page}\\\endfoot
\bottomrule\endlastfoot
age & Age in completed years; adult study restricted to 18 through 80. \\
duration\_days & Elapsed symptom duration at presentation, in days. \\
fever\_history\_gt38 & History of temperature strictly greater than 38 C during this illness. \\
fever\_past24h & Reported or measured fever during the preceding 24 hours. \\
temperature\_c & Measured temperature at presentation in degrees Celsius. \\
cough & Cough present during the current illness (0/\allowbreak{}1). \\
coryza & Coryza present during the current illness (0/\allowbreak{}1). \\
tonsillar\_exudate & Purulent tonsillar exudate/\allowbreak{}pus observed (0/\allowbreak{}1). \\
tonsillar\_swelling & Tonsillar swelling observed (0/\allowbreak{}1). \\
severe\_tonsillar\_inflammation & Severe tonsillar inflammation observed (0/\allowbreak{}1). \\
tender\_anterior\_cervical\_nodes & Tender anterior cervical lymphadenopathy/\allowbreak{}lymphadenitis observed (0/\allowbreak{}1). \\
household\_contact & Household exposure to a diagnosed GAS case (0/\allowbreak{}1). \\
\end{longtable}
\noindent All fields are generated observations, not actual clinical records. Field names match the frozen code; no downstream assay or outcome field is a predictor.
\par\endgroup
\renewcommand{\thefigure}{S\arabic{figure}}\renewcommand{\theHfigure}{S\arabic{figure}}\setcounter{figure}{0}
\clearpage
\begin{figure}[p]
\centering
\includegraphics[width=\linewidth,height=.70\textheight,keepaspectratio]{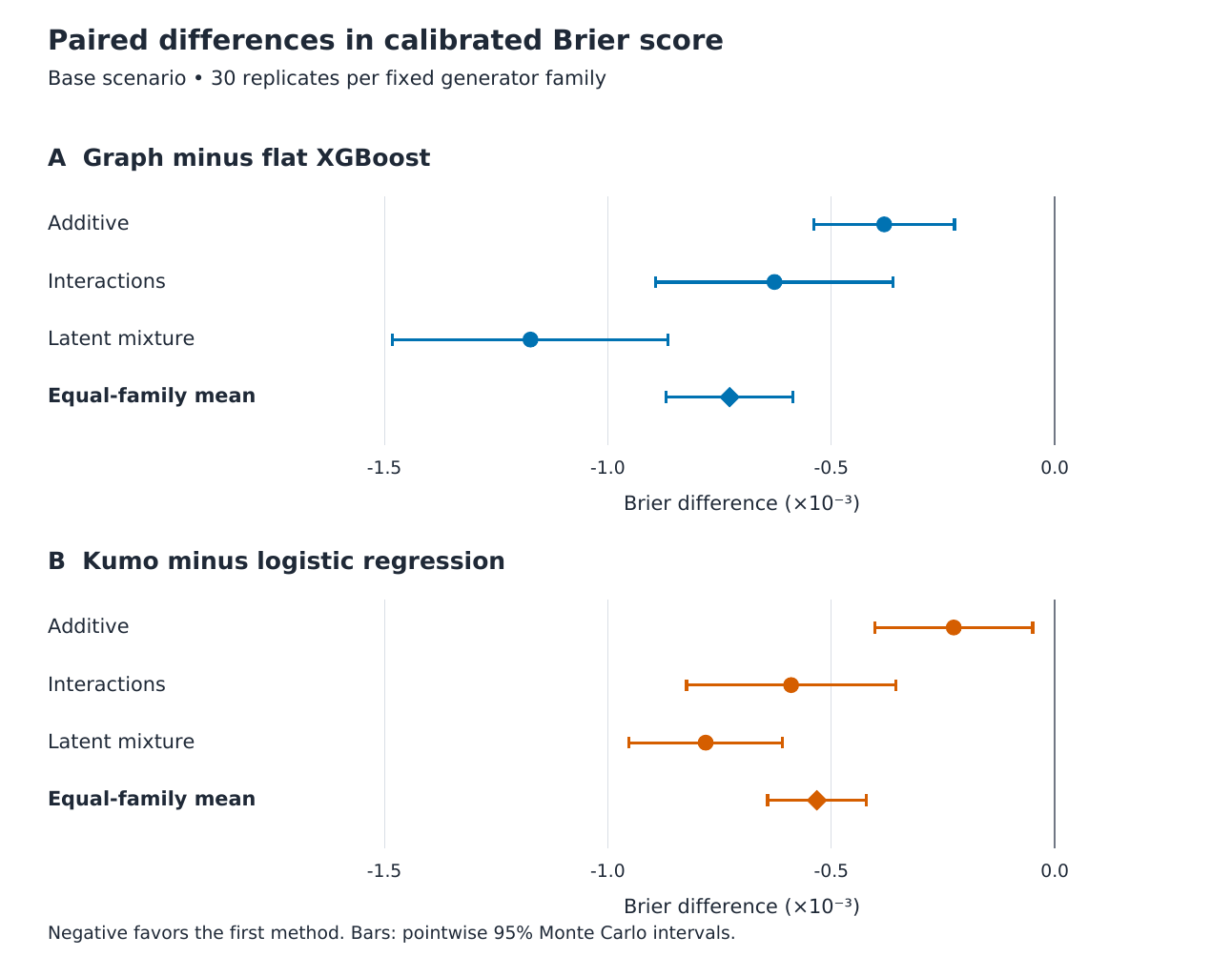}
\caption{\textbf{Paired differences in base-scenario calibrated Brier score}. Within-run Brier differences for graph versus flat XGBoost (A) and Kumo versus logistic regression (B). Negative differences favor the first method. Circles show family means from 30 independent replicate runs; diamonds show the mean with fixed equal weights across the three generator families (90 runs). Horizontal bars are pointwise 95\% Monte Carlo t intervals, using 29 degrees of freedom within family and Satterthwaite degrees of freedom for the equal-family mean. The graph-versus-flat comparison was the primary contrast; Kumo versus logistic was exploratory. Intervals describe simulation sampling error conditional on the specified worlds, are not multiplicity-adjusted, and do not quantify clinical uncertainty. Axis units are 0.001 Brier units.}\label{fig:support_Figure1}
\end{figure}
\clearpage
\clearpage
\begin{figure}[p]
\centering
\includegraphics[width=\linewidth,height=.70\textheight,keepaspectratio]{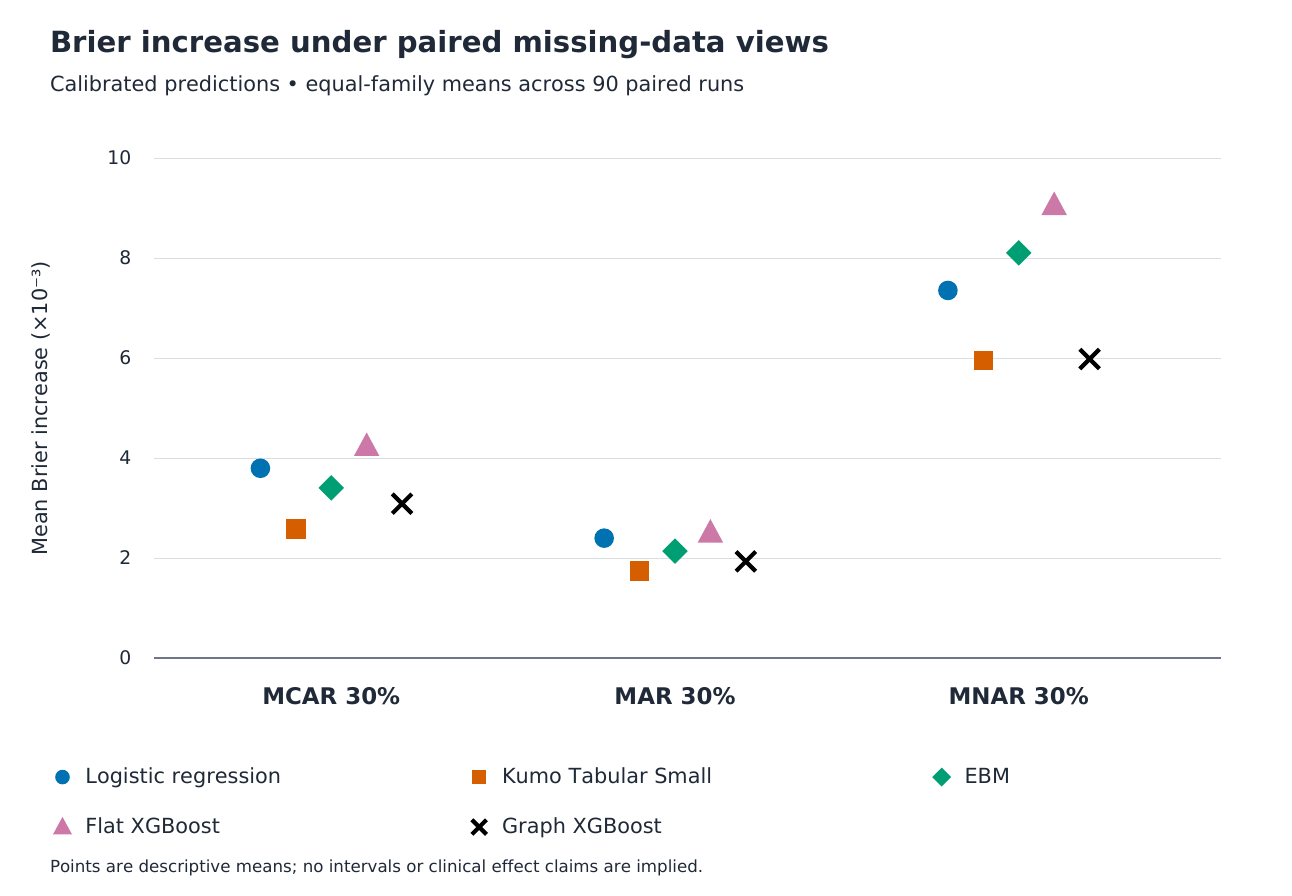}
\caption{\textbf{Change in Brier score under missing-data mechanisms}. For each calibrated method, the masked-view Brier score minus the base-view Brier score was computed within each run, then averaged with equal weights across the three fixed generator families. Each point summarizes 90 paired runs. MCAR, MAR and MNAR views reuse base encounters and outcomes. The nominal masking target is 30\% per eligible field (nine of twelve fields, about 22.5\% of all feature cells), compared with 10\% per eligible field in base data; realized rates vary. Greater positive values indicate greater loss of probability accuracy. Points show descriptive means without uncertainty bars. The horizontal offsets distinguish methods and have no numerical meaning. Axis units are 0.001 Brier units. These hypothetical missingness mechanisms do not establish robustness in clinical data.}\label{fig:support_Figure2}
\end{figure}
\clearpage
\clearpage
\begin{figure}[p]
\centering
\includegraphics[width=\linewidth,height=.70\textheight,keepaspectratio]{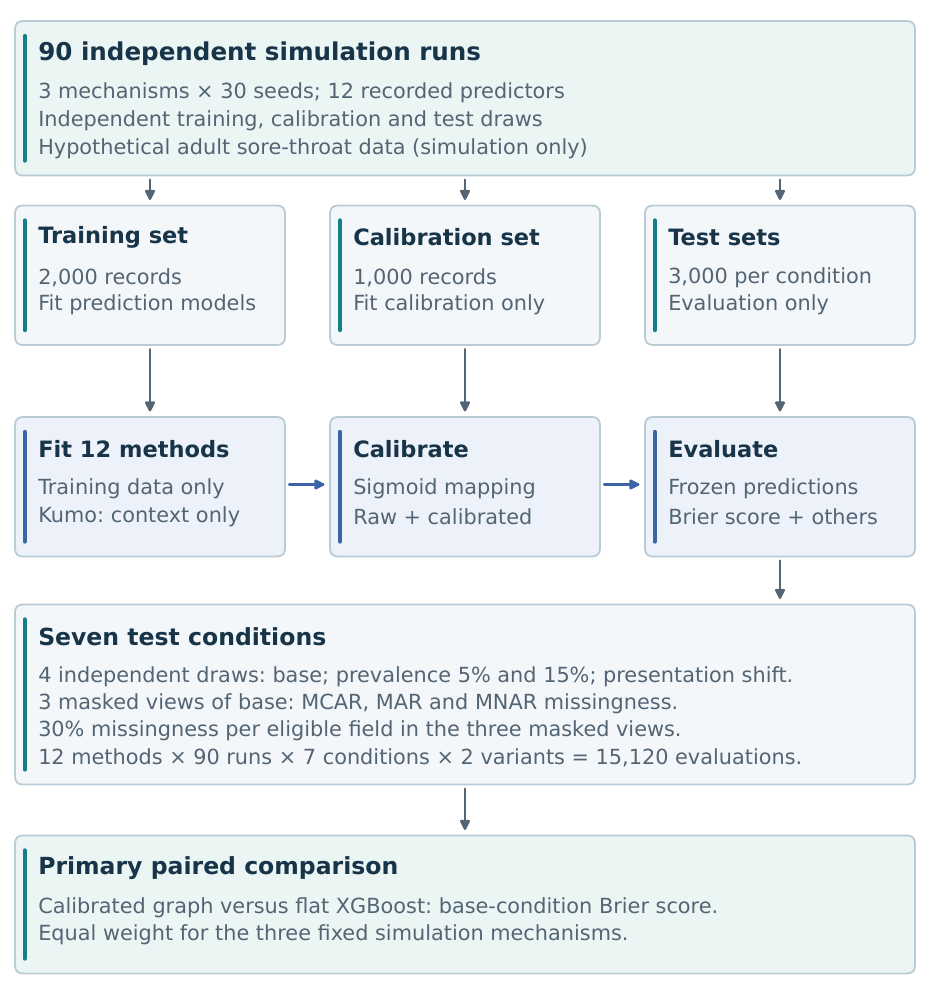}
\caption{\textbf{Supporting predictive-benchmark workflow}. Ninety simulations combine three fixed hypothetical mechanisms with 30 seeds each. Training, calibration and test data have separate roles. Three missingness stresses reuse underlying base test cases, yielding 1.35 million independent simulated encounters and 2.16 million exported rows across the full predictive benchmark. The communication extension uses only its base and three missingness views.}\label{fig:S3}
\end{figure}
\clearpage
\clearpage
\begin{figure}[p]
\centering
\includegraphics[width=\linewidth,height=.70\textheight,keepaspectratio]{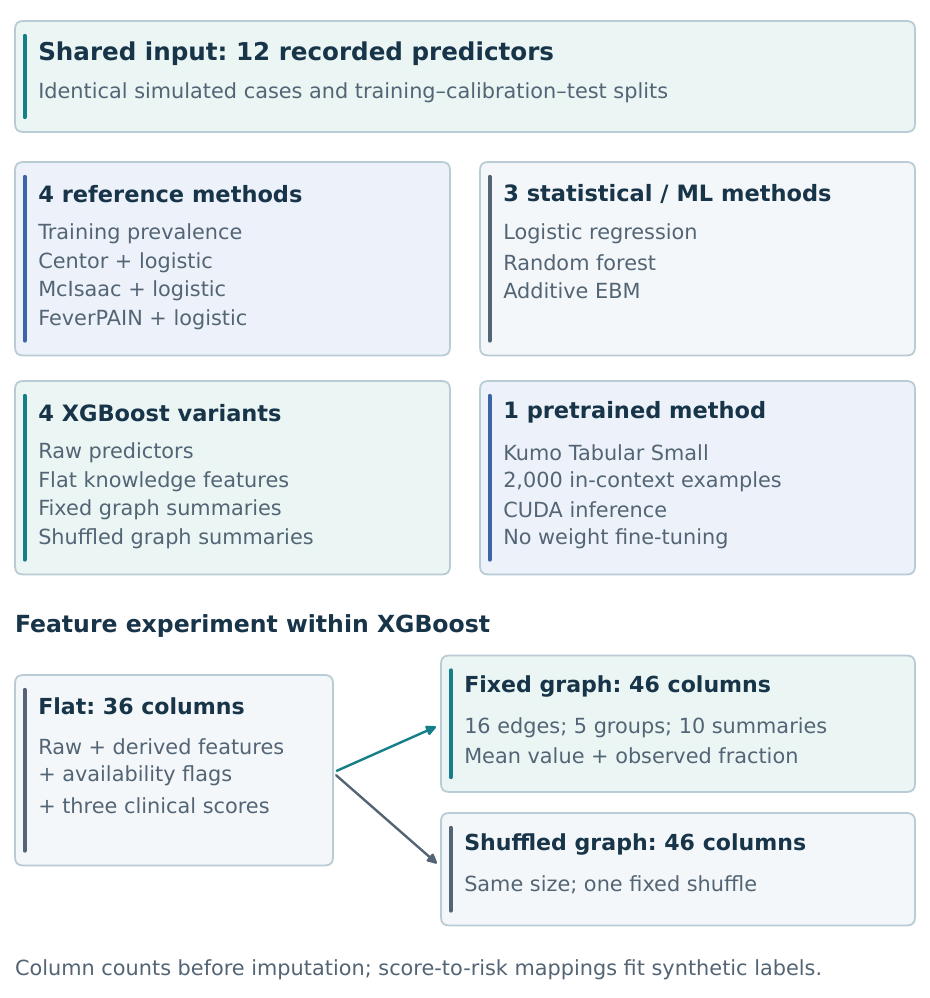}
\caption{\textbf{Supporting model comparisons and feature representations}. The twelve pipelines comprise four reference methods, three statistical or machine-learning models, four XGBoost representations and local Kumo Tabular Small. The clinical feature graph provides fixed summaries and is not a communication network or trained graph neural network. Five of these pipelines supply the initial opinions for the communication extension.}\label{fig:S4}
\end{figure}
\clearpage

\clearpage\begin{thebibliography}{99}
\bibitem{degroot} Morris H. DeGroot. Reaching a Consensus. \emph{Journal of the American Statistical Association}. 1974;69(345):118--121. \href{https://doi.org/10.1080/01621459.1974.10480137}{doi:10.1080/01621459.1974.10480137}.

\bibitem{boyd} Stephen Boyd, Arpita Ghosh, Balaji Prabhakar, and Devavrat Shah. Randomized Gossip Algorithms. \emph{IEEE Transactions on Information Theory}. 2006;52(6):2508--2530. \href{https://doi.org/10.1109/TIT.2006.874516}{doi:10.1109/TIT.2006.874516}.

\bibitem{ref1} Centers for Disease Control and Prevention. Clinical Guidance for Group A Streptococcal Pharyngitis. CDC clinical guidance; 2025. \href{https://www.cdc.gov/group-a-strep/hcp/clinical-guidance/strep-throat.html}{Official source}.

\bibitem{ref2} Centers for Disease Control and Prevention. Testing for Strep Throat or Scarlet Fever. CDC patient information; 2025. \href{https://www.cdc.gov/group-a-strep/testing/index.html}{Official source}.

\bibitem{ref3} National Institute for Health and Care Excellence. Sore throat (acute): antimicrobial prescribing (NG84): Terms used in the guideline. NICE guideline NG84; 2018. \href{https://www.nice.org.uk/guidance/ng84/chapter/terms-used-in-the-guideline}{Official source}.

\bibitem{ref4} Infectious Diseases Society of America. IDSA Clinical Practice Guideline Update on Group A Streptococcal (GAS) Pharyngitis. IDSA practice guideline; 2025. \href{https://www.idsociety.org/practice-guideline/streptococcal-pharyngitis2/}{Official source}.

\bibitem{ref5} Infectious Diseases Society of America. 2025 Clinical Practice Guideline Update by the Infectious Diseases Society of America on diagnosis and management of Group-A Streptococcal Pharyngitis (GAS): Risk assessment using clinical scoring systems in children and adults --- Supplemental Material. Supplement to IDSA practice guideline; 2025. \href{https://www.idsociety.org/globalassets/idsa/practice-guidelines/streptococcal-pharyngitits/gas-pharyngitis-pico-a-b-supplement.pdf}{Official source}.

\bibitem{ref6} Tim P. Morris; Ian R. White; Michael J. Crowther. Using simulation studies to evaluate statistical methods. Statistics in Medicine. 38(11):2074--2102; 2019. \href{https://doi.org/10.1002/sim.8086}{doi: 10.1002/\allowbreak{}sim.8086}.

\bibitem{ref7} Tianqi Chen; Carlos Guestrin. XGBoost: A Scalable Tree Boosting System. Proceedings of the 22nd ACM SIGKDD International Conference on Knowledge Discovery and Data Mining (KDD 2016); 2016. \href{https://doi.org/10.1145/2939672.2939785}{doi: 10.1145/\allowbreak{}2939672.2939785}.

\bibitem{ref8} InterpretML contributors. Explainable Boosting Machine. InterpretML official documentation; n.d. \href{https://interpret.ml/docs/ebm.html}{Official source}.

\bibitem{ref9} NVIDIA. Structured Data Models: SDM 0.2.1. GitHub software release; 2026. \href{https://github.com/NVIDIA/structured-data-models/releases/tag/v0.2.1}{Official source}.

\bibitem{ref10} Paul Little; Michael Moore; F. D. R. Hobbs; David Mant; Cliodna McNulty; Ian Williamson; Edith Cheng; Beth Stuart; Joanne Kelly; Jane Barnett; Mark Mullee; PRISM investigators. PRImary care Streptococcal Management (PRISM) study: identifying clinical variables associated with Lancefield group A \ensuremath{\beta}-haemolytic streptococci and Lancefield non-Group A streptococcal throat infections from two cohorts of patients presenting with an acute sore throat. BMJ Open. 3(10):e003943; 2013. \href{https://doi.org/10.1136/bmjopen-2013-003943}{doi: 10.1136/\allowbreak{}bmjopen-2013-003943}.

\bibitem{ref11} Harsha Nori; Samuel Jenkins; Paul Koch; Rich Caruana. InterpretML: A Unified Framework for Machine Learning Interpretability. arXiv:1909.09223; 2019. \href{https://doi.org/10.48550/arXiv.1909.09223}{doi: 10.48550/\allowbreak{}arXiv.1909.09223}.

\bibitem{ref12} NVIDIA. Kumo Tabular model card, revision v1.0.1. Hugging Face model repository; n.d. \href{https://huggingface.co/nvidia/Kumo-Tabular/blob/v1.0.1/README.md}{Official source}.

\end{thebibliography}
\end{document}